\documentclass[runningheads]{llncs}

\usepackage[T1]{fontenc}
\usepackage{cite}
\usepackage{amsmath,amssymb,amsfonts}
\usepackage{algorithmic}
\usepackage{graphicx}
\usepackage{textcomp}
\usepackage{xcolor}
\usepackage{comment}
\usepackage{amsmath}
\usepackage{booktabs}
\usepackage{todonotes}
\usepackage{listings}
\usepackage{enumitem}   
\usepackage{tabularx}
\usepackage{multirow}
\PassOptionsToPackage{hyphens}{url}
\usepackage[hidelinks]{hyperref}
\usepackage{textpos}
\usepackage{framed}

\usepackage[acronym]{glossaries}
\newacronym{anon_cred}{AC}{anonymous credentials}
\newacronym{cred_schema}{CS}{credential schema}

\newacronym{claim}{CL}{claim}
\newacronym{ver_claim}{VCL}{verifiable claim}
\newacronym{cred}{CR}{credential}
\newacronym{issuer_priv}{SK}{issuer private key}
\newacronym{issuer_pub}{PK}{issuer public key}
\newacronym{ver_cred}{VC}{verifiable credential}
\newacronym{cond_params}{CP}{conditional parameters}
\newacronym{input_pub}{In$_{pub}$}{public input}
\newacronym{input_priv}{In$_{priv}$}{private input}
\newacronym{proof_program}{PP}{proof program}
\newacronym{proving_key}{PP$_{key}$}{proving key}
\newacronym{ver_key}{VSC$_{key}$}{verification key}
\newacronym{offchain_comp}{VOC}{verf. off-chain computation}
\newacronym{ver_presentation}{VP}{verifiable presentation}
\newacronym{ver_presentations}{VPs}{verifiable presentations}
\newacronym{ver_presentation_out}{VP$_{out}$}{computation output}
\newacronym{ver_contract}{VSC}{verifier smart contract}

\newacronym{key_gen}{KeyGen}{KeyGen}
\newacronym{attest}{Attest}{Attest}
\newacronym{compile}{Compile}{Compile}
\newacronym{prove}{Prove}{Prove}
\newacronym{verify}{Verify}{Verify}

\newcommand{\short}[1]{\acrshort{#1}} 
\newcommand{\Short}[1]{\text{\acrshort{#1}}} 
\newcommand{\command}[1]{\texttt{\acrshort{#1}}} 

\usepackage{tikz}
\usepackage{pgfplots}
\usetikzlibrary{arrows,arrows.meta,positioning,calc,shapes,matrix,chains,decorations.pathreplacing,decorations.text,fit}
\usepackage{sansmath}

\tikzset{
  >=stealth',
  font={\sffamily\scriptsize},
  Entity/.style={fill=palette4!20},
}

\definecolor{palette1}{HTML}{0EB45A}
\definecolor{palette2}{HTML}{C51021}
\definecolor{palette3}{HTML}{009CDC}
\definecolor{palette4}{HTML}{999933}
\definecolor{palette5}{HTML}{DDCC77}
\definecolor{palette6}{HTML}{CC6677}
\definecolor{palette7}{HTML}{AA4499}
\definecolor{palette8}{HTML}{332288}

\pgfplotscreateplotcyclelist{mycolorlist}{%
palette1!50!black,every mark/.append style={fill=palette1},mark=*\\%
palette4!10!black,every mark/.append style={fill=palette4},mark=square*\\%
palette7!50!black,every mark/.append style={fill=palette7},mark=otimes*\\%
black,mark=star\\%
blue,every mark/.append style={fill=blue!80!black},mark=diamond*\\%
red,densely dashed,every mark/.append style={solid,fill=red!80!black},mark=*\\%
brown!60!black,densely dashed,every mark/.append style={solid,fill=brown!80!black},mark=square*\\%
black,densely dashed,every mark/.append style={solid,fill=gray},mark=otimes*\\%
blue,densely dashed,mark=star,every mark/.append style=solid\\%
red,densely dashed,every mark/.append style={solid,fill=red!80!black},mark=diamond*\\%
}

\pgfplotscreateplotcyclelist{mycolorlistbarplots}{%
palette1!50!black,every mark/.append style={fill=palette1},fill=palette1,mark=none\\%
palette4!10!black,every mark/.append style={fill=palette4},fill=palette4,mark=none\\%
palette7!50!black,every mark/.append style={fill=palette7},fill=palette7,mark=none\\%
}

\pgfplotscreateplotcyclelist{mycolorlistbarplots1}{%
palette4!10!black,every mark/.append style={fill=palette4},fill=palette4,mark=none\\%
palette7!50!black,every mark/.append style={fill=palette7},fill=palette7,mark=none\\%
}

\pgfplotscreateplotcyclelist{mycolorlistbarplots2}{%
palette1!50!black,every mark/.append style={fill=palette1},fill=palette1,mark=none\\%
}

\pgfplotsset{
	legend style={font=\tiny\sffamily},
	label style={font=\footnotesize\sffamily},
	tick label style={font=\footnotesize\sffamily},
	ticklabel style={font=\footnotesize\sffamily},
	xticklabel style={font=\footnotesize\sffamily},
	yticklabel style={font=\footnotesize\sffamily},
    discard if not/.style 2 args={
        x filter/.code={
            \edef\tempa{\thisrow{#1}}
            \edef\tempb{#2}
            \ifx\tempa\tempb
            \else
                
            \fi
        }
    }
}

\pgfplotsset{compat=1.11,
    /pgfplots/ybar legend/.style={
    /pgfplots/legend image code/.code={%
       \draw[##1,/tikz/.cd,yshift=-0.25em]
        (0cm,0cm) rectangle (3pt,0.8em);},
   },
}

\newcommand{\bettershortstack}[2][c]{
  \begin{tabular}[b]{@{}#1@{}}#2\end{tabular}
}

\makeatletter
\pgfdeclareshape{document}{
\inheritsavedanchors[from=rectangle] 
\inheritanchorborder[from=rectangle]
\inheritanchor[from=rectangle]{center}
\inheritanchor[from=rectangle]{north}
\inheritanchor[from=rectangle]{south}
\inheritanchor[from=rectangle]{west}
\inheritanchor[from=rectangle]{east}
\backgroundpath{
\southwest \pgf@xa=\pgf@x \pgf@ya=\pgf@y
\northeast \pgf@xb=\pgf@x \pgf@yb=\pgf@y
\pgf@xc=\pgf@xb \advance\pgf@xc by-10pt 
\pgf@yc=\pgf@yb \advance\pgf@yc by-10pt
\pgfpathmoveto{\pgfpoint{\pgf@xa}{\pgf@ya}}
\pgfpathlineto{\pgfpoint{\pgf@xa}{\pgf@yb}}
\pgfpathlineto{\pgfpoint{\pgf@xc}{\pgf@yb}}
\pgfpathlineto{\pgfpoint{\pgf@xb}{\pgf@yc}}
\pgfpathlineto{\pgfpoint{\pgf@xb}{\pgf@ya}}
\pgfpathclose
\pgfpathmoveto{\pgfpoint{\pgf@xc}{\pgf@yb}}
\pgfpathlineto{\pgfpoint{\pgf@xc}{\pgf@yc}}
\pgfpathlineto{\pgfpoint{\pgf@xb}{\pgf@yc}}
\pgfpathlineto{\pgfpoint{\pgf@xc}{\pgf@yc}}
}
}
\makeatother

\makeatletter
\pgfdeclareshape{artifact}{
\inheritsavedanchors[from=rectangle] 
\inheritanchorborder[from=rectangle]
\inheritanchor[from=rectangle]{center}
\inheritanchor[from=rectangle]{north}
\inheritanchor[from=rectangle]{south}
\inheritanchor[from=rectangle]{west}
\inheritanchor[from=rectangle]{east}
\backgroundpath{
\southwest \pgf@xa=\pgf@x \pgf@ya=\pgf@y
\northeast \pgf@xb=\pgf@x \pgf@yb=\pgf@y
\pgf@xc=\pgf@xb \advance\pgf@xc by-2pt
\pgf@yc=\pgf@ya \advance\pgf@yc by+2pt
\pgfpathmoveto{\pgfpoint{\pgf@xa}{\pgf@ya}}
\pgfpathlineto{\pgfpoint{\pgf@xa}{\pgf@yb}}
\pgfpathlineto{\pgfpoint{\pgf@xb}{\pgf@yb}}
\pgfpathlineto{\pgfpoint{\pgf@xb}{\pgf@yc}}
\pgfpathlineto{\pgfpoint{\pgf@xc}{\pgf@ya}}
\pgfpathclose
\pgfpathmoveto{\pgfpoint{\pgf@xc}{\pgf@ya}}
\pgfpathlineto{\pgfpoint{\pgf@xc}{\pgf@yc}}
\pgfpathlineto{\pgfpoint{\pgf@xb}{\pgf@yc}}
\pgfpathlineto{\pgfpoint{\pgf@xc}{\pgf@yc}}
}
}
\makeatother

\begin{document}

\title{Non-Disclosing Credential On-chaining for Blockchain-based Decentralized Applications}
\titlerunning{Non-Disclosing Credential On-chaining for DApps}

\author{Anonymous Submission}
\authorrunning{Anonymous Submission}
\institute{}

\author{Jonathan Heiss\inst{1} \and
Robert Muth\inst{2} \and
Frank Pallas\inst{1} \and
Stefan Tai\inst{1}}

\authorrunning{J. Heiss et al.}

\institute{Information Systems Engineering, Technische Universität, Berlin, Germany \and
Distributed Security Infrastructures, Technische Universität, Berlin, Germany \\
\email{\{j.heiss,muth,frank.pallas,tai\}@tu-berlin.de}
}

\maketitle

\begin{textblock*}{\textwidth}(0cm,16cm) 
\begin{center}
\begin{framed}
    \textit{Preprint to appear in the Proceedings of the \textbf{20\textsuperscript{th} International Conference on Service-Oriented Computing (ICSOC 2022)}.}
\end{framed}
\end{center}
\end{textblock*}


\begin{abstract}
Many service systems rely on verifiable identity-related information of their users. Manipulation
and unwanted exposure of this privacy-relevant information, however, must at the same time be prevented and avoided.
Peer-to-peer blockchain-based decentralization with a smart contract-based execution model and verifiable off-chain computations leveraging zero-knowledge proofs promise to provide the basis for next-generation, non-disclosing credential management solutions.
In this paper, we propose a novel credential on-chaining system that ensures blockchain-based transparency while preserving pseudonymity.
We present a general model compliant to the W3C verifiable credential recommendation and demonstrate how it can be applied to solve existing problems that require computational identity-related attribute verification.
Our zkSNARKs-based reference implementation and 
evaluation show that, compared to related approaches based on, e.g., CL-signatures, our approach provides significant performance advantages and more flexible proof mechanisms, underpinning our vision of increasingly decentralized, transparent, and trustworthy service systems.
\end{abstract}


\sloppy

\section{Introduction}
\label{sec:introduction}
Blockchain-based Decentralized Applications~(DApps) are service systems where the backend code runs on a peer-to-peer blockchain network, using smart contracts for the application logic. DApps are increasingly applied
in distrusted, multi-stakeholder environments
to overcome reliance and dependence on
trusted and often centralized third parties~(TTPs) and associated risks of failure, manipulation, or opportunistic behavior.
Traditional architectures are transitioned into decentralized ones
in that core functionalities provided by previously centralized TTPs are now implemented
in smart contracts, which are executed on the blockchain by each peer~\cite{DApps_SoK_2021}.
This way, involved stakeholders do not have to trust TTPs anymore to act as intended, resulting in a service system that is more transparent and manipulation-resistant.

Even when using permissionless blockchains, a DApp may still define roles, permissions, or assets and assign them to particular users and other parties.
DApps therefore depend on reliably distinguishing the users by some kind of distinct \emph{attributes} that only a single participant or a group of users possesses and for which a \emph{proof} can be provided.
In decentralized finance (DeFi) lending DApps~\cite{SOK_DeFi_Londoo}, for instance, lenders and borrowers may be required to prove possession of a valid citizenship or tax number, in IoT DApps~\cite{blogpv_peise_2021}, devices may have to prove official calibration or certain configuration parameters, and participants of marketplaces~\cite{service_marketplaces_klems} may be required to prove creditworthiness before they are allowed to engage into trading.
The attestation of such attributes can come from different \emph{issuers} such as tax offices, calibration authorities, or credit bureaus.
Parties having to prove attributes can then do so by presenting \emph{credentials} allowing another party to \emph{verify} the fulfillment of attribute requirements.

In state-of-the-art service systems, DApps which need to verify user attributes typically employ some off-chain authority for this purpose acting as a TTP and providing respective services.
For example, DApps running on permissioned blockchains rely on identity and access management~(IAM) services that are provided at platform-level through a trusted committee of nodes, e.g., the Membership Service Providers authority in Hyperledger Fabric~\cite{misc:hyperledger}. 
However, in many DApps, off-chain verification of user attributes is not appropriate as it violates design goals. 
DeFi and marketplace DApps, for example, implement cryptocurrency-based transaction logic that may depend on the on-chain verification of user attributes.
In IoT DApps that typically characterize through uncertain, dynamic, and distrusted settings, it may simply not be possible to employ an off-chain TTP service for authentication that is tamper-resistant and always online.
Furthermore, many DApps call for independent verifiability of all transactions, including attribute verification, e.g., for external auditing.

Alternatively, credential verification can be implemented as smart contract-based logic as well and be executed as part of the blockchain’s consensus protocol~\cite{SmartContracts_AnonCred_Muth_2022}.
This mitigates the problems described above, however, it also introduces other challenges originating from the blockchain's natural design, most importantly related to users' privacy:
While users of DApps running on permissionless blockchains can veil their identity behind pseudonymous account addresses, on-chain credential verification threatens this pseudonymity since consensus-based validation reveals confidential user attributes to the blockchain network. 
Existing approaches that, for example, leverage \emph{zero-knowledge proofs}~(ZKP) to keep such attributes off the blockchain, either suffer from severe performance limitations making their usage impractical~\cite{SmartContracts_AnonCred_Muth_2022} or are limited to a specific use case and credential type which restricts general applicability~\cite{zokVehicles_florida_2019,ZokCarSharing_TUDresden_2020,zokHealthcare_indian_2020}.

In face of the on-going debate about measures for Know-Your-Customer~(KYC) and Anti-Money Laundering~(AML), and consequential blockchain regulations\footnote{\url{https://www.sec.gov/news/statement/crenshaw-defi-20211109}}\footnote{\url{https://www.europarl.europa.eu/news/en/press-room/20220627IPR33919}}, we consider the need for on-chain credential verification as real and technical approaches essential to pave the way for general purpose adoption of DApps as well as presenting an alternative to using traditional off-chain IAM services. 
To this end, we herein provide an approach to make attribute-specific user credentials verifiable on the blockchain while preserving pseudonymity properties. 
We, thereby, make three individual contributions: 
\begin{itemize}
    \item First, we propose a novel credential on-chaining system. This system leverages \emph{verifiable off-chain computations}~(\short{offchain_comp})~\cite{off-chaining_models_heiss} for executing logic on confidential identity attributes through the holder and for only presenting a non-revealing ZKP to the peer network.
    \item Second, we show how the proposed model can be employed to solve typical computational identity-related problems by proposing different types of credential conditions to be verified. 
    \item Third, we demonstrate technical feasibility by providing a reference implementation for each type of condition using ZoKrates~\cite{zokrates_eberhardt}, a tool for realizing VOC on Ethereum~\cite{ethereum:yellowpaper}. Our implementation exhibits significant performance advantages over on-chain verification of established \emph{Camenisch-Lysyanskaya}~(CL) signature-based ZKPs~\cite{SmartContracts_AnonCred_Muth_2022}. 
\end{itemize}

In the remainder of this paper, we first describe relevant concepts and related work in Section~\ref{sec:preliminaries}.
Then, we present a general system design for credential on-chaining in Section~\ref{sec:systemDesign}.
On this basis, in Section~\ref{sec:application} we show how to apply the design to realize different proof types.
Details on the implementation and an evaluation based on our proofs-of-concept~(PoC) are described in Section~\ref{sec:Evaluation}.
Finally, we conclude with some final remarks in Section~\ref{sec:Conclusion}.

\section{Preliminaries}
\label{sec:preliminaries}
As relevant preliminaries, we first introduce the core idea of on-chaining credentials underlying our approach along with its benefits and challenges.
Then, we describe existing concepts and approaches that are relevant to our contributions.

\subsection{On-chaining Verifiable Credentials}
Within this paper, on-chaining verifiable credentials describes the process of providing credentials originating from off-chain to smart contracts in a verifiable manner.
Instead of having a blockchain-external entity checking the validity of an attribute-based credential (such as holding a particular citizenship or being creditworthy) and acting as a centralized TTP service provider, the issuer-generated credential is verified on-chain as a smart contract-based transaction that is validated through the blockchain's consensus protocol by each peer in the network. 

\textbf{Benefits:} For DApps, on-chain credential verification has some considerable benefits over off-chain verification:
\begin{itemize}
    \item \textit{Transparency:} Credentials are independently verifiable throughout and, in public blockchains, even beyond the network.
    
    \item \textit{Tamper-resistance:} While off-chain authorities could previously manipulate credential verification unnoticedly, on-chain verification prevents this. 

    \item \textit{Passive verification:} The credential verification is self-executed and automatically verified by all blockchain peers to maintain global consensus~\cite{whoAmI_2017}.
    
    \item \textit{Availability:} It is not required to have an off-chain authority to be online all the time. Availability is guaranteed by the blockchain network.
    
    \item \textit{Immediate usage:} The output of the credential verification can immediately be used on-chain as part of the smart contract-based application logic. 

    \end{itemize}

\textbf{Challenges:} On-chain credential verification also introduces challenges that originate from the nature of blockchains.
\begin{itemize}
    \item \textit{Privacy:} Given that the blockchain-based system is a fully replicated one with full transparency and an immutable, append-only data structure, confidential information contained in credentials must not be revealed on-chain.

    \item \textit{Verification costs:} Given fully redundant transaction execution as part of expensive consensus protocols, e.g., Proof-of-Work, the computational costs for verifying credentials on-chain should be kept at a minimum.

    \item \textit{On-/Off-chain interactions:} Given an isolated execution environment that restricts interactivity with off-chain systems, the DApp cannot directly call the issuer to check, e.g., credential authenticity. Therefore, credentials must become non-interactively verifiable and function without trusted oracles~\cite{trustworthyOnchaining_heiss}.

\end{itemize}

\subsection{Related Work and Concepts}
Our contributions build upon the concepts of \emph{verifiable credentials}~(\short{ver_cred}),
\emph{anonymous credentials}~(\short{anon_cred}), and \emph{verifiable off-chain computations}~(\short{offchain_comp}), and intersect with related work around these existing concepts.

\begin{figure}[b] %
	\centering
	\begin{tikzpicture}[
	auto,
	node distance=2em,
	entity/.style={draw, rectangle, rounded corners=.25em, minimum height=2em, minimum width=5.8em},
	cornered/.style={draw, artifact, minimum height=1.5em, minimum width=2em}
]
	\node[entity](issuer) {Issuer};
	\node[cornered, right=of issuer](ver_cred) {VC};
	\node[entity, right=of ver_cred](holder) {Holder};
	\node[cornered, right=of holder](ver_presentation) {VP};
	\node[entity, right=of ver_presentation](verifier) {Verifier};
	
	\draw[rectangle, rounded corners=.25em] ($(issuer.south west) - (0, 1em)$) rectangle node[anchor=center] {Verifiable Data Registry~(VDR)} ($(verifier.south east) - (0, 3em)$);
	
	\node[cornered, below=1.25em of issuer, xshift=-1.5em](cred_schema) {CS};
	\node[right=0pt of cred_schema] {(...)};
	
	\path[draw, -] (issuer) -- (ver_cred);
	\path[draw, ->] (ver_cred) -- (holder);
	\path[draw, -] (holder) -- (ver_presentation);
	\path[draw, ->] (ver_presentation) -- (verifier);
	
	\path[draw, <->] (issuer.south) -- ++(0, -1em);
	\path[draw, <->] (holder.south) -- ++(0, -1em);
	\path[draw, <->] (verifier.south) -- ++(0, -1em);
	
	\path[draw, ->, dashed] (verifier) edge [bend right=1.75em] node[yshift=.15em] {Trust} (issuer); 
\end{tikzpicture} 
	\caption{W3C Verifiable Credential Model}
	\label{fig:vc_model}
\end{figure}
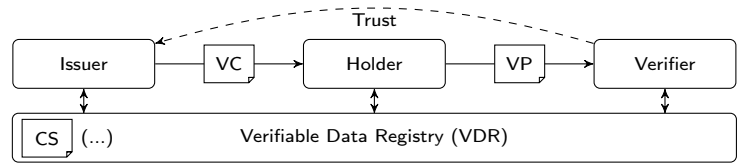

\textbf{Verifiable Credentials: }
\label{subsub:VC}
The W3C recommendation for \short{ver_cred}s~\cite{misc:W3_VerifCred} advocates a user-centric identity management model where claims on identity attributes are assumed to be attested by a trusted \emph{issuer} and issued as \short{ver_cred} to the \emph{holder} where they are securely stored, e.g., in a wallet.
The \emph{holder} can then independently present a selection of verifiable attribute claims as a \emph{\acrlong{ver_presentation}}~(\short{ver_presentation}) to a \emph{verifier}, as illustrated in Figure~\ref{fig:vc_model}.
Different from the objective of this paper, the VC model assumes the verifier to be executed off the blockchain.
Blockchains are applied only to realize \emph{Verifiable Data Registries}~(VDR) that are used to store public artifacts, such as identifiers, public keys, or \emph{credential schemas}~(CS) which describe what \short{ver_cred} consist of and how they are verified.

Blockchain-based implementations of this model include credential management systems such as uPort~\cite{uPort_naik_2020}, Jolocom~\cite{jolocom:whitepaper}, or Hyperledger Indy~\cite{misc:HLIndyWalkthrough}. None of these does, however, implement on-chain credential verification.

\textbf{Anonymous Credentials: }
\short{anon_cred} have been proposed to enable the verification of credentials without revealing confidential user attributes to the verifier.
A ZKP is generated to convince a verifier about certain aspects of the credential.
Common proof types for credentials include \emph{range proofs} to verify that a value is within a given range or \emph{set membership proofs} to verify that an element is part of a predefined set~\cite{DBLP:conf/asiacrypt/CamenischCS08}.

CL signature-based ZKPs~\cite{CLSig} are a well-known approach to \short{anon_cred} that have been implemented, for example, in IBM’s attribute-based credential system, Identity Mixer~\cite{idemix_2002}, which, in turn, has been adopted in Hyperledger Indy~\cite{misc:HLIndyWalkthrough}. 
As another zero-knowledge protocol class that distinguishes through non-interactivity and succinct proof size, zkSNARKs~(zero-knowledge succinct non-interactive argument of knowledge) have also been applied to enable non-revealing credential verification~\cite{ZKlaims_2019}.

\textbf{Verifiable Off-chain Computation}
Another essential concept for on-chain credential verification is \short{offchain_comp}~\cite{off-chaining_models_heiss} which has been introduced to mitigate blockchain's limitations regarding privacy and scalability. 
Here, the result of any off-chain computation can be verified on-chain without revealing private inputs to the computation.
\short{offchain_comp} has conceptually be extended in~\cite{heiss_trustworthypreprocessing_ICSOC2021} where it is considered as an intermediate pre-processing step in data on-chaining workflows between an off-chain data source and an on-chain verifier. 
To technically realize \short{offchain_comp}, ZoKrates~\cite{zokrates_eberhardt} has been proposed, a toolbox and language for the construction of on-chain verifiable ZKPs based on zkSNARKs.

ZoKrates has been adopted in various use-cases to enable non-revealing authentication of DApp users. 
Examples include smart vehicle authentication at charging stations~\cite{zokVehicles_florida_2019}, user authentication for car sharing~\cite{ZokCarSharing_TUDresden_2020}, or patient authentication in health care~\cite{zokHealthcare_indian_2020}.
These works so far do, however, focus on the verification of a specific identity attribute only, consider comparably trivial authentication schemes, or lack general applicability.
A rather general approach to on-chain credential verification is described in~\cite{SmartContracts_AnonCred_Muth_2022}. Here, it has been shown how CL signature-based ZKPs can be verified on-chain, albeit with considerable performance limitations currently rendering the approach impractical. 


\section{System Design}
\label{sec:systemDesign}
Seizing on the previously described challenges and limitations of existing approaches, in this section, we present our credential on-chaining system that applies \short{offchain_comp}~\cite{off-chaining_models_heiss} as pre-processing step~\cite{heiss_trustworthypreprocessing_ICSOC2021} to the W3C recommendations for \short{ver_cred}s~\cite{misc:W3_VerifCred}.
Instead of verifying the issuer-generated \short{ver_cred} directly on-chain, it is pre-processed by the user as a \short{offchain_comp} that returns a ZKP which can be verified by the DApp in a non-disclosing manner.

\begin{figure}[t] %
	\centering
	\input{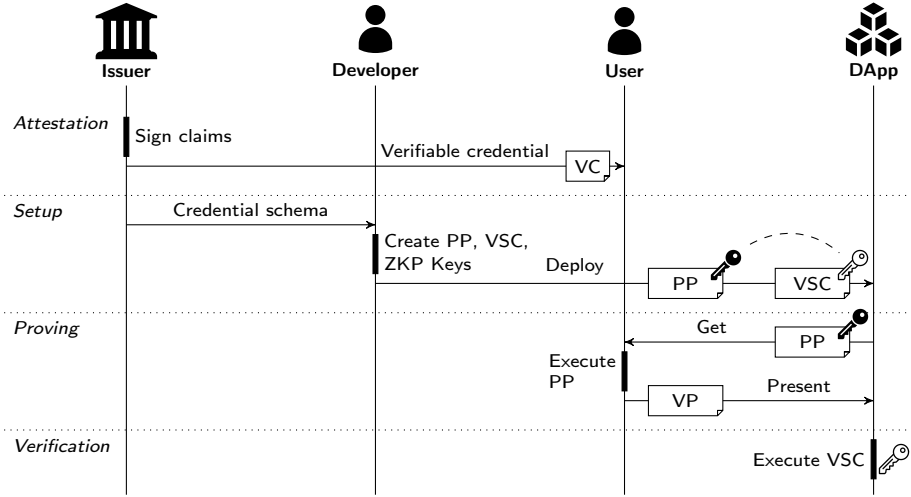}
	\caption{Credential on-chaining system overview which demonstrates all steps from initial signing credentials to the verification of a corresponding verifiable presentation}
	\label{fig:workflow}
\end{figure}

As illustrated in Figure~\ref{fig:workflow}, our system works along four stages, each of them executed by a different system role.

\begin{enumerate}

    \item During \textbf{attestation} which is considered a pre-requisite of our system the issuer signs identity claims contained in a credential and sends them as \short{ver_cred} to the user. 

    \item The \textbf{setup} describes all activities executed by the developer to create the artifacts required for proving and verification, i.e., the \emph{\acrlong{proof_program}}~(\short{proof_program}) that implements the verification logic as a \short{offchain_comp}, the \emph{ZKP keys} used to sign and verify the \acrlong{ver_presentation}~(\short{ver_presentation}), and the \emph{\acrlong{ver_contract}}~(\short{ver_contract}) that is required by the DApp to verify the \short{ver_presentation}.
    
    \item During \textbf{proving}, a DApp user creates a \short{ver_presentation} from her \short{ver_cred} using the \short{proof_program} and the \emph{proving key}~(\short{proving_key}). 
    
    \item The \textbf{verification} of the \short{ver_presentation} consisting of a ZKP and a computational result is executed on-chain by the \short{ver_contract} using the verification key~(\short{ver_key}).
\end{enumerate}


\subsection{Attestation}
During the attestation, the issuer creates a \short{ver_cred} from a \emph{credential}~(\short{cred}) according to a public credential schema: $\command{attest}(\Short{cred}, \Short{issuer_priv}) \xrightarrow{} (\Short{ver_cred})$.
A \short{cred} consists of a set of \emph{claims}~(\Short{claim}) such that $\Short{cred} = \{\Short{claim}_1, ..., \Short{claim}_n\}$.
A claim, in turn, consists of a 3-tuple comprising subject, attribute, and value, e.g., $(\text{Alice}, \text{age}, \text{31})$.
The issuer signs the credential with its individual \emph{issuer secret key} $\Short{issuer_priv}$ which can be verified with the corresponding \emph{issuer public key} $\Short{issuer_pub}$.
To enable users to select single claims from a \short{ver_cred} (and, hence, enable \emph{selective disclosure}) the issuer needs to sign each claim of a credential individually instead of all claims collectively.
This results in a \Short{ver_cred} consisting of a set of \emph{verifiable claims}~(\Short{ver_claim}) such that $\Short{ver_cred} = \{\Short{ver_claim}_1,...,\Short{ver_claim}_n\}$.
A \Short{ver_claim} in turn consists of a claim-signature pair: $\Short{ver_claim} = \{\short{claim},SIG\}$.
Once attested through the issuer, the \short{ver_cred} is stored by the holder in her personal wallet. 
While VCs of different issuers may be used, for simplicity, we describe the following stages assuming a single issuer. 

\subsection{Setup}
During the setup, the developer creates required artifacts which enable the construction and verification of ZKPs. 
Therefore, she obtains the public credential schema from the issuer (or indirectly from a public VDR as described in Section~\ref{subsub:VC}).
Without compromising generality of our approach, we consider a zkSNARKs-based setup that takes the technological capabilities of ZoKrates~\cite{zokrates_eberhardt} into account
and consists of the following three steps:

First, the verification logic and required input types are implemented for an execution environment that represents the \short{proof_program} and enables the assertion of computational correctness through a ZKP. 
Using ZoKrates, its high-level language can be leveraged for this purpose which compiles into an executable constraint system represented in the ZoKrates Intermediate Representation format~\cite{zokrates_eberhardt}. 

Second, the developer generates the \emph{ZKP keys} from the $\Short{proof_program}: \command{key_gen}(\Short{proof_program})\xrightarrow{}(\Short{proving_key}, \Short{ver_key})$.
The ZKP keys are bound to the \Short{proof_program} and enable a prover, here the user, to create a verifiable \Short{proof_program}-specific ZKP with the proving key~(\short{proving_key}) and a verifier, here the DApp, to verify the ZKP with the corresponding verification key~(\short{ver_key}).

As a third step, the developer implements the \short{ver_presentations}-verification logic in the \emph{verifier smart contracts}~(\short{ver_contract}), 
integrates all artifacts, i.e., the \Short{proof_program}, the \Short{issuer_pub}, the \Short{ver_contract}, the \Short{ver_key}, 
into the DApp and deploys it. 
While the deployed DApp only requires the \Short{ver_contract} and the \Short{ver_key} for the verification, it makes the \Short{proof_program} and \Short{issuer_pub}, which are required for proving, accessible to the users. 

\begin{figure}[t] %
	\centering
	\input{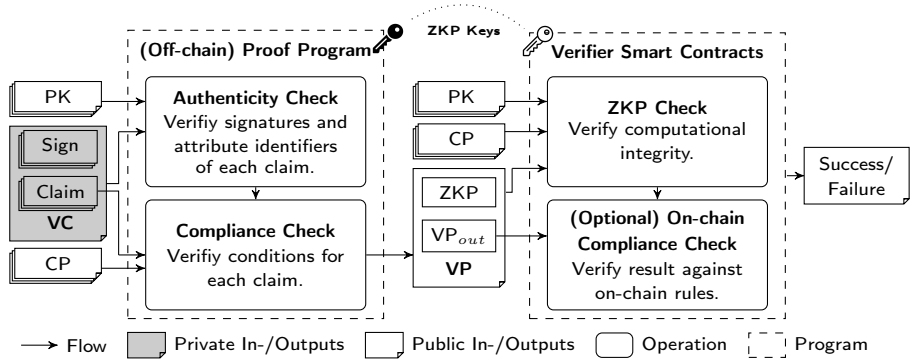}
	\caption{In- and outputs of the proof program and the verifier smart contracts}
	\label{fig:PP_VP}
\end{figure}

\subsection{Proving}
\label{design_proving}
In the proving stage, the user first obtains the \short{proof_program} and \Short{proving_key} from the DApp and selects the required \Short{ver_claim}s from her personal wallet. 
Based on that, she executes the proving:
$\command{prove}(\Short{proving_key}, \Short{input_pub}, \Short{input_priv})\xrightarrow{} (\text{ZKP}, \Short{ver_presentation_out})$.
The \short{proof_program} takes two types of inputs: \emph{public inputs}~(\Short{input_pub}) are required on-chain for ZKP verification and are, hence, revealed to the blockchain network whereas \emph{private inputs}~(\Short{input_priv}) are kept off-chain.
To keep confidential attributes hidden, \Short{ver_claim}s are treated as \Short{input_priv}, but \emph{conditional parameters}~(\short{cond_params}) and \Short{issuer_pub}s that need to be reviewed on-chain are treated as \Short{input_pub}.
As depicted in Figure~\ref{fig:PP_VP}, the \short{proof_program} executes two checks on each verifiable claim contained in a \short{ver_cred}:
\begin{enumerate}[label=(\roman*)]
    \item \textit{Authenticity Check}: To guarantee that a verifiable claim has been attested to by the right issuer (trusted by the developer), its signature is verified with the issuer's \Short{issuer_pub}: $F_{auth} \colon (\Short{ver_claim}, \Short{issuer_pub}) \xrightarrow[]{} (bool)$. 
    Furthermore, the integrity of the attribute is reviewed, i.e., that the user provides the correct attribute to the \short{proof_program}, by comparing the provided attribute identifier contained in the claim against an attribute identifier predefined by the developer. 
    \item \textit{Compliance Check}: To prove compliance with context- and credential-specific conditions that the developer defines based on the DApp logic, the \Short{claim}'s attribute value is checked against some \short{cond_params}, e.g., age higher than 21: $F_{comp} \colon (\Short{claim}, \Short{cond_params}) \xrightarrow[]{} (bool)$ 
\end{enumerate}

The output is a \Short{ver_presentation} that contains a ZKP for the correct \short{proof_program} execution and the corresponding \emph{computional output}~(\Short{ver_presentation_out}), e.g., a boolean value indicating if the Authenticity and Compliance Checks succeed.
The \Short{ver_presentation} does not contain \Short{input_priv} anymore; hence, it can be presented without risking the \short{ver_cred}s confidentiality.

\subsection{Verification}
The verification is executed on-chain through the verifier smart contracts~(\short{ver_contract}s): $\command{verify}(\Short{ver_key}, \text{ZKP}, \Short{ver_presentation_out}, \Short{cond_params}, \Short{issuer_pub}) \xrightarrow{} (bool)$.
Inputs to the \short{ver_contract}s are the \Short{ver_presentation} consisting of the ZKP and the \Short{ver_presentation_out} and the public inputs used for proving, i.e., the conditional parameters \Short{cond_params} and the issuers' public key \Short{issuer_pub}.
As depicted in Figure~\ref{fig:PP_VP}, the \short{ver_contract}s implement two checks: 
\begin{enumerate}[label=(\roman*)]
    \item \emph{ZKP Check}: To verify that the proving has correctly been executed on the expected public inputs, the ZKP Check is executed on the VP and the public inputs: $F_{zkp} \colon (\text{ZKP}, \Short{ver_presentation_out}, \Short{issuer_pub}_{i},\Short{cond_params}) \xrightarrow[]{}  (bool)$.
    \item \emph{Compliance Check}: Optionally, an additional on-chain Compliance Check is executed on \Short{ver_presentation_out} as, for example, required for the Uniqueness Proof mechanism presented in Section~\ref{sub:uniqueness}.
\end{enumerate}

\section{Application}
\label{sec:application}
Given the proposed credential on-chaining system, in this section, we demonstrate the system's proving abilities through a set of proof mechanisms.
On the one hand, we show how established concepts, e.g., range and membership proofs~\cite{DBLP:conf/asiacrypt/CamenischCS08}, can be realized with our system, on the other hand, we introduce novel mechanisms to on-chain credential verification in DApps, namely, relative time-dependent proofs and uniqueness proofs.

To describe the proof mechanisms, we set a particular focus on the off- and on-chain Compliance Checks (CompCheck) and the conditional parameters (CP) used to validate credential-specific conditions.
In contrast to the off-chain Authenticity Check (AuthCheck) and the on-chain ZKP Check which are conceptionally application-agnostic, the CompChecks are application-specific.

\subsection{Range and Equality Proofs}
\label{sub:range_proofs}
\emph{A DApp requires a numeric attribute to be in a specific range indicated through an upper and/or lower bound.}
This may, for example, be required in referendum DApps~\cite{bbb}, to guarantee that only DApp users in a specific zip code range are eligible to participate.

\textbf{Off-chain Proving: } 
For range proofs, the CompCheck validates if the attribute value is within the range defined by the developer through one or two boundaries.
Given a Turing-complete language as with ZoKrates, range proofs can be implemented as simple predicated statements.
\short{cond_params}s are range boundaries that are provided as public inputs to the \short{proof_program}.
Thereby, the same \short{proof_program} can be used for different ranges simply by setting different range boundaries as public inputs.

\textbf{On-chain Verification: } 
Since the range boundaries are required as public inputs for the ZKP Check, it can independently be verified that the range has been set correctly. 
No further on-chain CompCheck is required.

\textbf{Discussion: }
A range proof will not reveal identity attributes to the verifier as they are defined as private inputs, but if the boundaries of a range proof are too small, the private attribute can be approximated or even completely exposed.
It is also possible that past proofs with different boundaries can be combined, so that the intersections of all proofs reveal insights of the \short{ver_cred} or even the exact attribute value.
This is especially dangerous with a publicly available transaction history, e.g., in permissionless blockchains.
To avoid this, proofs should not be linkable to each other or with a single blockchain account.

A special type of range is the \emph{equality proof} which can be implemented with the described range proof mechanism.
Here, the range is set to a single value that needs to match the user's attribute value.
While we consider equality proofs with private attributes pointless since a successful verification discloses the attribute on-chain, DApps may have reasons to require clear text attributes from a user.
However, once published, attributes are publicly revealed and cannot be removed due to the permanent nature of a blockchain, so we advise the greatest attention to privacy for such proofs.
Also, it needs to be mentioned, that while equality proofs can easily be realized with our system, there are more efficient approaches, e.g., by verifying the issuer's signature directly on-chain.

\subsection{Relative Time-dependent Proofs}
\label{sub:time_proofs}
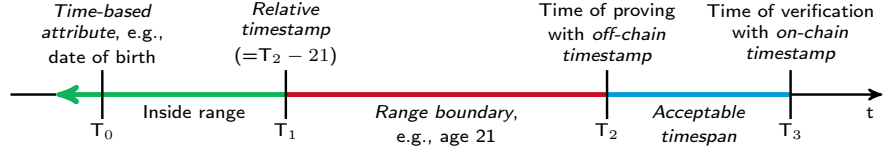
\begin{figure}[h] %
	\centering
	\begin{tikzpicture}[
		tik/.style={yshift=2.5em}
	]
		\foreach \y [evaluate=\y as \yy using \y * (\textwidth / 20)] in {0,...,19} {
			\coordinate (p-\y) at (\yy pt, 0);
		}
		
		\path[draw, -,  thick, line width=.1em] (p-0) -- (p-1);
		\path[draw, <-, thick, line width=.2em, palette1] (p-1) -- (p-2);
		\path[draw, -,  thick, line width=.2em, palette1] (p-2) -- node[below, black] {Inside range} (p-6);
		\path[draw, -,  thick, line width=.2em, palette2] (p-6) -- node[below, black] {\shortstack[c]{\emph{Range boundary},\\e.g., age 21}}(p-13);
		\path[draw, -,  thick, line width=.2em, palette3] (p-13) -- node[below, black] {\shortstack[c]{\emph{Acceptable}\\\emph{timespan}}} (p-17);
		\path[draw, ->, thick, line width=.1em] (p-17) -- (p-19);
		
		\foreach \x [count=\i from 0] in {2, 6, 13, 17} {
			\path[draw, -,  thick, line width=.1em] ([yshift=1em]p-\x) -- ([yshift=-1em]p-\x);
			\node[yshift=-1.5em] at (p-\x) {$\text{T}_\i$};
		}
		
		\node[yshift=-0.75em, xshift=-.5em] at (p-19) {$\text{t}$};
	
		\node[tik] at (p-2) {\shortstack[c]{\emph{Time-based}\\\emph{attribute}, e.g.,\\date of birth}};
		\node[tik] at (p-6) {\shortstack[c]{\emph{Relative}\\\emph{timestamp}\\(=$\text{T}_2 - 21$)}};
		\node[tik] at (p-13) {\shortstack[c]{Time of proving\\with \emph{off-chain}\\\emph{timestamp}}};
		\node[tik] at (p-17) {\shortstack[c]{Time of verification\\with \emph{on-chain}\\\emph{timestamp}}};
	\end{tikzpicture}
	
	\caption{Age verification scenario for time-dependent proof mechanisms}
	\label{fig:time-proof}
\end{figure}

\emph{A DApp requires a date- or time-based attribute to be in a range that has boundaries relative to the current date or timestamp.}
This is, for example, required if the users' age needs to be checked or if an expiration date of a credential needs to be verified. 
Realizing such proofs with our system is challenging in that a timestamp is required off-chain for ZKP generation, however, the off-chain timestamp is not verified on-chain as part of the consensus protocol and, hence, can be manipulated unnoticedly.
To describe our approach to that, we assume that a DApp needs to verify that the age of its users is above 21, i.e., a range proof with a single boundary, and the required time-dependent user attribute is the date of birth (cf. Figure~\ref{fig:time-proof}).

\textbf{Off-chain Proving: } 
\short{cond_params}s are the \emph{range boundary}, e.g., age~21, that is predefined by the developer, and the \emph{relative timestamp} that is pre-calculated by the user as the current date minus the range boundary, i.e., $2022 - 21 = 2001$.
For simplicity, timestamps are here indicated in years, although more fine-grained timestamps are applicable.
Both, the relative timestamp and the range boundary, are provided to the \Short{proof_program} as public inputs.
As part of the off-chain CompCheck, the \Short{proof_program} compares the relative timestamp to the date of birth that is part of the \Short{ver_cred} and, hence, provided as private inputs to the \Short{proof_program}.
If the relative timestamp is larger than the date of birth, the off-chain CompCheck is successful. 

\textbf{On-chain Verification: } 
On-chain, an additional CompCheck is required to make sure that the off-chain timestamp has not been faked by the user.
Therefore, the DApp takes the on-chain timestamp from the current block header which can be considered trusted as it is validated through the blockchain’s consensus protocol, and compares it against the off-chain timestamp.
If on- and off-chain timestamp are within an \emph{acceptable timespan} pre-determined by the developer the on-chain CompCheck succeeds.

\textbf{Discussion:} 
Relative time-dependent proofs are broadly applicable, especially considering their usage for validating expiration dates of credentials, e.g., for driver's licenses or credit cards.
In this case, the expiration date needs to be signed by the issuer together with the corresponding attribute(s) to make sure that both belong together. 
However, it must also be noted that only rough periods can be verified with the proposed mechanism.
For example, the acceptable time span between proving and verification is strongly dependent on peculiarities of the applied blockchain's consensus protocol, e.g., block interval and confirmation time, which may vary in orders of several minutes.

\subsection{Set Membership Proofs}
\emph{A DApp requires that a holder's attribute value $val$ is in a predefined finite set $S = \{s_1, s_2, ..., s_n\}$ such that $val \in S$.}
Such proofs can be used to show that a holder belongs to a set of authorized users in permissioned settings as, for example, required in referendum DApps where eligible participants are, a priori, equipped with a referendum voucher. 
To prove set membership without revealing $val$ on-chain, we leverage path proofs in Merkle Trees similar to the one described here\footnote{\url{https://github.com/Zokrates/ZoKrates/tree/deploy/zokrates_cli/examples/merkleTree}}.



\textbf{Off-chain Proving: }
In addition to the user's set attribute contained in the set that is provided as private input, the \Short{proof_program} requires three \Short{cond_params}s that need to be pre-computed by the holder and are provided as public inputs to the \Short{proof_program}: (1) the root hash of the Merkle Tree constructed over $S$, (2) an array of hash values situated on the path from the leaf hash (hashed attribute), to the root hash, and (3) a same-size array of left-right indicators that determine in which order two child hashes are concatenated as the pre-image of the parent hash.
During the \Short{proof_program}'s CompCheck, first the attribute's hash is calculated. 
Then, the PP iterates over both, the hash and indicator arrays, and, in each iteration, calculates the next upper hash in the Merkle Tree.
If the resulting hash is equal to the root hash provided as public input, set membership is proven.

\textbf{On-chain Verification: }
On-chain, correctness of the \Short{proof_program} computation and its public inputs is reviewed during the ZKP Check. 
No further on-chain CompCheck is required. 

\textbf{Discussion: } 
Different types of set membership proofs are well-known and have extensively been discussed in the literature~\cite{DBLP:conf/crypto/Merkle87, DBLP:conf/fc/BenarrochCFGK21}.
We consider the set entries to be confidential such that they must not be revealed on the blockchain. 
An access list for DApp users generated off-chain, for example, must not reveal the user identifiers, e.g., to prevent attackers to simply use them for registration.
In the proposed mechanism, only hashes of entries are revealed on-chain but the pre-images which are required for successful ZKP generation remain off-chain, kept secret by the user.
It should be noted that privacy guarantees become stronger with an increasing size of set $S$, although the verification complexity grows only logarithmically due to the tree data structure.

In some cases, set membership proofs can be used interchangeably with range proofs.
If DApp users need to prove that they live in a specific city without revealing it, a set membership proof can be leveraged that builds upon city identifiers in a given state or a range proof can be constructed based on the zip code applicable for the city.

\subsection{Uniqueness Proof}
\label{sub:uniqueness}
\emph{A DApp requires a unique pseudonymous identifier~(UPI), to distinguish different users.}
This allows DApps to prevent Sybil-attacks~\cite{sybil_attacks_doceur} and to establish user accountability, e.g., function calls can be mapped to users even if different account addresses are used. 
Such UPIs are, for example, required if user-specific access control policies need to be verified as part of the DApp.
However, with respect to the user privacy, the UPI must provide unlinkability (1) to off-chain activities of the user, i.e., a relation to the \emph{real} off-chain identity must not be revealed, and (2) to on-chain activities of the user, i.e., UPIs of the same user used in different DApps must not be linkable with each other. 

For this mechanism, we assume that a user can uniquely be identified through the issuer, either through a single unique identifier, e.g., a tax identification number, or through a set of attributes that jointly enable unique identification.  
Latter is assumed in the following.
Furthermore, we assume that a unique DApp identifier is available, e.g., the address of the verifier smart contract.

\textbf{Off-chain Proving: }
The set of predefined attributes is provided to the \Short{proof_program} by the user as private inputs and individually validated as part of the AuthCheck. 
As a single conditional parameter, the unique DApp identifier is provided as public input.
The UPI construction is executed as part of the CompCheck: The attribute values and the DApp identifier are concatenated in a predefined order and the resulting concatenation is hashed representing the UPI: $\texttt{Hash}(att_1, ..., att_n, \textit{ID}_{\textit{DApp}})$. 
Together with the ZKP, the UPI is returned as \Short{ver_presentation}.

\textbf{On-chain Verification: }
On receiving the \Short{ver_presentation}, first, the ZKP is checked together with all public inputs, i.e., the DApp identifier and the public keys. 
A successful validation attests correct construction of the UPI, but unique user registration has not yet been proven. 
Therefore, the DApp checks during the on-chain CompCheck if the UPI has already been registered in a \emph{user registry} that is maintained by the DApp and contains the UPIs and account addresses of all registered users. 
If the UPI is already part of the list, the CompCheck fails. 
Otherwise, the UPI is recorded to the list and the user is officially registered. 

\textbf{Discussion: }
The proposed mechanism provides unlinkability to off- and on-chain activities of users: since the preimage of the hash is a concatenation of both, user attributes and the DApp identifier, the attributes applied for UPI construction cannot be traced back and the UPI changes for each DApp that a user registers at. 
This provides for user privacy and makes the mechanism applicable to various DApp contexts where uniqueness is required, e.g., votings, token airdrops, and access control.

However, the mechanism only provides uniqueness of users if the previous assumptions hold. 
If the attributes change, the user can create a different UPI and use it for registering twice with different blockchain accounts, which eliminates Sybil-resistance. 


\section{Evaluation}
\label{sec:Evaluation}
To evaluate our credential on-chaining system and its applicability to the proposed proof mechanisms, in this section, we first provide technical details for the ZoKrates-based implementations of the proof mechanisms, then analyze performance aspects, particularly in comparison with a similar CL-signature-based approach, and finally discuss open issues and refinements.

\subsection{Implementation}
We implement each proof mechanisms prototypically to demonstrate the technical feasibility of our proposal and provide the source codes on GitHub\footnote{\url{https://github.com/JonathanHeiss/ZoKrates-Credential-Verification}}. 

For \emph{attestation} we provide a Python script which constructs a EDDSA signature on a test credential -- employing a Python library for ZoKrates-compatible cryptographic instructions\footnote{\url{https://github.com/Zokrates/pycrypto}} -- and returns a \Short{ver_cred}. 
For the \emph{setup}, we implement the \Short{proof_program} using the ZoKrates language which can be compiled using the ZoKrates command line interface~(CLI) which also enables the creation of the ZKP keys (i.e., \Short{proving_key} and \Short{ver_key}) and the \acrlong{ver_contract}~\Short{ver_contract}. 
Once the artifacts are created, the \emph{proving} is executed in two CLI-aided steps:
First, a \emph{witness} is generated with the command \texttt{compute-witness} which represents an input-specific variable assignment of the executable constraint system.
Second, the ZKP is created based on the witness and the proving key with the command \texttt{generate-proof}.
For \emph{verification}, a Solidity \Short{ver_contract} that is automatically generated by means of the ZoKrates toolbox implements the routines required to verify the ZKP in the \texttt{verytx()} function using the integrated \Short{ver_key}.
The optional on-chain Compliance Check is implemented in a separate smart contract.

To execute these processes we provide scripts that also measure the execution times, the artifact sizes, and the verification costs with the latter being enabled through a Truffle\footnote{\url{https://trufflesuite.com}} test project on a simulated Ethereum~\cite{ethereum:yellowpaper} blockchain.

\subsection{Performance Analysis}

\begin{table}[t]
\centering
\caption{Gas costs, artifact sizes, and execution times for proof generation and verification with ZoKrates.}
\begin{tabular}{lrrrrrrr}
\toprule
    \bf Proof & 
    \bf \hspace{-.4em}\bettershortstack[r]{TX Cost\\(Gas)} & 
    \bf \hspace{-.4em}\bettershortstack[r]{Witness\\(s)} & 
    \bf \hspace{-.4em}\bettershortstack[r]{Setup\\(s)} & 
    \bf \hspace{-.4em}\bettershortstack[r]{Proof\\(s)} & 
    \bf \hspace{-.4em}\bettershortstack[r]{Compiled\\(MB)} & 
    \bf \hspace{-.4em}\bettershortstack[r]{\short{proving_key}\\(MB)} & 
    \bf \hspace{-.4em}\bettershortstack[r]{\short{ver_key}\\(KB)} \\
\midrule

Equality & 511\,k & 4 & 63 & 13 & 352 & 38 & 8 \\
Range & 532\,k & 3 & 62 & 14 & 352 & 38 & 8 \\
Uniqueness & 675\,k & 8 & 89 & 27 & 864 & 47 & 12 \\
Rel. Time & 627\,k & 4 & 61 & 14 & 352 & 38 & 8 \\
Membership & 532\,k & 6 & 147 & 26 & 656 & 90 & 8 \\

\bottomrule
\end{tabular}
\label{table:measurements}
\end{table}

Based on our ZoKrates-based reference implementations,
we execute an experimental test case
for each proof mechanism
using test credentials that consist of a single claim,
such that, the Authenticity and Compliance check are only executed once.
All experiments were executed on a MacBook Pro (13", 2018) with a 2.3\,GHz Quad-Core Intel i5 and 8\,GB memory.

\textbf{ZoKrates-based Credential On-chaining Proof Mechanisms}
The measurements of our ZoKrates proof mechanism test cases
are depicted in Table~\ref{table:measurements}
which shows the corresponding 
blockchain transaction costs in Gas\footnote{Gas is an Ethereum-specific metric for measuring blockchain transaction complexity.},
execution runtimes in seconds, and artifact file sizes in MB.

The \emph{transaction costs} range between 511\,k and 675\,k~Gas. 
The costs are higher if more public inputs are passed to the \Short{proof_program}, e.g., compared to the equality proof, the range proof additionally requires \Short{cond_params}s.
Furthermore, if an additional compliance check is executed as, for example, for uniqueness and relative time-dependent proofs, transaction costs increase as well.

\emph{Execution times} are measured for the ZKP setup,
which in ZoKrates is required for generating the ZKP keys, 
and the proving, which consists of the witness and proof generation.
Most time-intensive is the setup which, however, is only executed once by the developer and, hence, does not impact the operation of the system where different users submit \Short{ver_presentations} to the DApp.
For operations, the proof generation time takes the longest, however, the longest proving time does only slightly exceed half a minute considering witness and proof generation together.
Regarding the different proving mechanisms, it can, as expected, be observed that with increasing computational complexity, the execution times increase as well.
An in-depth analysis about the behavior of ZoKrates for varying inputs and computations can be found in~\cite{eberhardt_PhD_thesis}.

The \emph{artifacts size} varies between artifact types, e.g., the compiled \Short{proof_program} size lies between 352 and 864\,MB and the \Short{proving_key} size between 38 and 90\,MB, whereas the \Short{ver_key} size is considerably smaller, ranging between 8 and 12\,KB.
This makes the latter suitable to be used on-chain where resources are scarce.
Consequences of large artifacts regarding the system deployment are discussed in Section~\ref{sub:discussion}.

\textbf{Comparison to CL-Signatures}
Furthermore, to establish comparability of performance behavior beyond our own proof mechanisms,
we implement and conduct three test cases that mirror the experimental evaluation of smart contract-based verification with CL-signatures as described in~\cite{SmartContracts_AnonCred_Muth_2022}.
These test cases are executed on multiple attributes and implement
(1) an equality proof,
(2) a range proof,
and (3) a combination of both.

In comparison,
as depicted in Table~\ref{table:zokratesVsCL},
with our approach, we are able to reduce the transactions costs in Gas by a factor of
$\approx 50$ to $100$ in all three test cases.
This distinct performance improvement shows that currently, in face of expensive on-chain computations, zkSNARKs are better suited for \emph{on-chain} credential verification than CL-signatures-based approaches.

\begin{table}[t]
\centering
\caption{Gas costs for on-chain proof verification with the same test credentials as in~\cite{SmartContracts_AnonCred_Muth_2022} for our implementation approach and with on-chain CL-signature verification.}

\begin{tabular}{p{7em}rp{8em}}
\toprule
\textbf{Proof} & \centering \textbf{ZoKrates} & \hspace{1.2em}\textbf{CL-Signature} \\

\midrule

Test: Equality & \raggedleft 593\,k & \hfill 32,001\,k \\ 
Test: Range & \raggedleft 521\,k & \hfill 84,826\,k \\ 
Test: Combined & \raggedleft 870\,k & \hfill 84,033\,k \\ 

\bottomrule
\end{tabular}
\label{table:zokratesVsCL}
\end{table}

\subsection{Discussion}
\label{sub:discussion}
\textbf{Deployment: }
Referring to Table~\ref{table:measurements}, we consider the size of the \Short{proof_program} and the proving key as too large to be stored on the blockchain. 
Consequently, provisioning of both artifacts to the user needs to be solved differently. 
As one solution, we propose to apply a content addressable storage pattern~\cite{off-chain_patterns_eberhardt} as for example realizable with IPFS and adopted for managing off-chain assets associated with Non-fungible Tokens~(NFTs).
Instead of storing both artifacts on-chain, only the artifacts' hash is stored on-chain and serves as a pointer to the artifacts that are stored off-chain in IPFS.
By comparing the on-chain hash-based address with the hash of the off-chain stored artifacts, integrity can be reviewed.
However, availability is not guaranteed per se~\cite{DBLP:journals/comsur/DanielT22}.
Therefore, an additional protocol, e.g., Filecoin, could be applied that introduces redundant storage and provides incentives for keeping off-chain files available.

\textbf{Revocation: }
A credential may be required to be revoked by the issuer, e.g., if a driver license is invalidated by a public authority.
Since revocation is not in the scope of this work, we propose to leverage dedicated blockchain-based revocation systems as an extension to our credential on-chaining system similar to the ones proposed for educational credentials in~\cite{credential_revocation_2020}.
Also, in some cases, revocation can be replaced with expiration dates on credentials which can be realized with the relative time-dependent proofs (see Section~\ref{sub:time_proofs}).

\textbf{Attacks: }
For identification purposes, especially for secret credentials, \emph{replay attacks} render a severe problem on blockchains.
Since proofs become available to everyone on the blockchain after the on-chain verification, they can be used by anybody else on the blockchain to fake a different identity or to gain unauthorized access.
To address this problem, developers can leverage the uniqueness proofs that enable the DApp to identify repeated submissions of the same proof from different users.
Also, we recommend to use a proving scheme other than Groth16~\cite{groth16_jensGroth_2016} which is vulnerable to \emph{malleability attacks}, or to implement countermeasures.

\section{Conclusion}
\label{sec:Conclusion}
How can identity-related attribute information and computations on such information be shared in both a fully transparent but pseudonymity-preserving manner? This question is not easy to answer, especially in service systems that comprise multiple autonomous service providers and consumers for which no mutual trust assumptions can and should be made. State-of-the-art, typically centralized trusted third parties providing IAM services, stop short when it comes to the trust assumptions expected and the associated risks of manipulation or opportunistic behavior existing.

In this paper, we present a novel, non-disclosing credential management system that builds on peer-to-peer decentralization through blockchain, smart contract-based execution, and \short{offchain_comp} using ZKPs.
We introduce a W3C recommendation-compliant \short{ver_cred} model, demonstrate and discuss typical identity-related computational problems that can now be addressed in such settings, and show technical feasibility through reference implementations with a cost and performance assessment.
The significant transaction cost improvements towards a comparable approach based on established CL-signatures underline the relevance of our approach for its application in practical settings.

Our work is in line with prior work on data on-chaining 
and contributes to the larger question on what data and what computation to handle on-chain, that is, on the blockchain when a blockchain is part of a larger service system, and what data and what computation to handle off-chain -- that is, anywhere but the blockchain -- while ensuring and not violating key system properties related to performance, security, trustworthiness, and other system quality criteria.

\Urlmuskip=0mu plus 1mu
\def\UrlBreaks{\do\/\do-}
\bibliographystyle{splncs04}
\bibliography{references}

\begin{thebibliography}{10}
\providecommand{\url}[1]{\texttt{#1}}
\providecommand{\urlprefix}{URL }
\providecommand{\doi}[1]{https://doi.org/#1}

\bibitem{whoAmI_2017}
Azouvi, S., Al-Bassam, M., Meiklejohn, S.: Who am i? secure identity
  registration on distributed ledgers. In: Data Privacy Management,
  Cryptocurrencies and Blockchain Technology. Springer International Publishing
  (2017)

\bibitem{DBLP:conf/fc/BenarrochCFGK21}
Benarroch, D., Campanelli, M., Fiore, D., Gurkan, K., Kolonelos, D.:
  Zero-knowledge proofs for set membership: Efficient, succinct, modular. In:
  Financial Cryptography. LNCS, Springer (2021)

\bibitem{DBLP:conf/asiacrypt/CamenischCS08}
Camenisch, J., Chaabouni, R., Shelat, A.: Efficient protocols for set
  membership and range proofs. In: {ASIACRYPT}. LNCS, Springer (2008)

\bibitem{CLSig}
Camenisch, J., Lysyanskaya, A.: A signature scheme with efficient protocols.
  In: {SCN}. LNCS, Springer (2002)

\bibitem{idemix_2002}
Camenisch, J., Van~Herreweghen, E.: Design and implementation of the idemix
  anonymous credential system. In: Proceedings of the 9th ACM Conference on
  Computer and Communications Security (2002)

\bibitem{DBLP:journals/comsur/DanielT22}
Daniel, E., Tschorsch, F.: {IPFS} and friends: {A} qualitative comparison of
  next generation peer-to-peer data networks. {IEEE} Commun. Surv. Tutorials
  (2022)

\bibitem{sybil_attacks_doceur}
Douceur, J.R.: The sybil attack. In: {IPTPS}. LNCS, Springer (2002)

\bibitem{eberhardt_PhD_thesis}
Eberhardt, J.: Scalable and privacy-preserving off-chain computations. Doctoral
  thesis, Technische Universität Berlin (2021)

\bibitem{off-chaining_models_heiss}
Eberhardt, J., Heiss, J.: Off-chaining models and approaches to off-chain
  computations. In: Proceedings of the 2Nd Workshop on Scalable and Resilient
  Infrastructures for Distributed Ledgers. SERIAL'18, ACM (2018)

\bibitem{off-chain_patterns_eberhardt}
Eberhardt, J., Tai, S.: On or off the blockchain? insights on off-chaining
  computation and data. In: {ESOCC}. LNCS, Springer (2017)

\bibitem{zokrates_eberhardt}
Eberhardt, J., Tai, S.: {ZoKrates} - scalable privacy-preserving off-chain
  computations. In: IEEE International Conference on Internet of Things
  (iThings) and IEEE Green Computing and Communications (GreenCom) and IEEE
  Cyber, Physical and Social Computing (CPSCom) and IEEE Smart Data (SmartData)
  (2018)

\bibitem{zokVehicles_florida_2019}
Gabay, D., Akkaya, K., Cebe, M.: A privacy framework for charging connected
  electric vehicles using blockchain and zero knowledge proofs. In: IEEE 44th
  LCN Symposium on Emerging Topics in Networking. pp. 66--73 (2019)

\bibitem{groth16_jensGroth_2016}
Groth, J.: {On the Size of Pairing-Based Non-Interactive Arguments}. In: 35th
  Annual International Conference on Advances in Cryptology. LNCS, Springer
  (2016)

\bibitem{ZokCarSharing_TUDresden_2020}
Gudymenko, I., Khalid, A., Siddiqui, H., Idrees, M., Clauß, S., Luckow, A.,
  Bolsinger, M., Miehle, D.: Privacy-preserving blockchain-based systems for
  car sharing leveraging zero-knowledge protocols. In: IEEE International
  Conference on Decentralized Applications and Infrastructures (DAPPS) (2020)

\bibitem{heiss_trustworthypreprocessing_ICSOC2021}
Heiss, J., Busse, A., Tai, S.: {Trustworthy Pre-Processing of Sensor Data in
  Data On-chaining Workflows for Blockchain-based IoT Applications}. In: 19th
  International Conference on Service-Oriented Computing. LNCS, Springer (2021)

\bibitem{trustworthyOnchaining_heiss}
Heiss, J., Eberhardt, J., Tai, S.: From oracles to trustworthy data on-chaining
  systems. In: IEEE International Conference on Blockchain (2019)

\bibitem{misc:HLIndyWalkthrough}
{Hyperledger Indy-SDK Repository}: Indy walkthrough - a developer guide for
  building indy clients using libindy (2018),
  \url{https://github.com/hyperledger/indy-sdk/blob/master/docs/getting-started/indy-walkthrough.md}

\bibitem{misc:hyperledger}
{Hyperledger White Paper Working Group}: An introduction to hyperledger (2018),
  \url{https://www.hyperledger.org/wp-content/uploads/2018/07/HL\_Whitepaper\_IntroductiontoHyperledger.pdf}

\bibitem{jolocom:whitepaper}
{JOLOCOM}: A decentralized, open source solution for digital identity and
  access management (whitepaper) (2019),
  \url{https://jolocom.io/wp-content/uploads/2019/12/Jolocom-Whitepaper-v2.1-A-Decentralized-Open-Source-Solution-for-Digital-Identity-and-Access-Management.pdf}

\bibitem{service_marketplaces_klems}
Klems, M., Eberhardt, J., Tai, S., H{\"a}rtlein, S., Buchholz, S., Tidjani, A.:
  Trustless intermediation in blockchain-based decentralized service
  marketplaces. In: Service-Oriented Computing. Springer International
  Publishing (2017)

\bibitem{DBLP:conf/crypto/Merkle87}
Merkle, R.C.: A digital signature based on a conventional encryption function.
  In: {CRYPTO}. LNCS, Springer (1987)

\bibitem{bbb}
Muth, R., Eisenhut, K., Rabe, J., Tschorsch, F.: {BBBlockchain}:
  Blockchain-based participation in urban development. In: eScience. {IEEE}
  (2019)

\bibitem{SmartContracts_AnonCred_Muth_2022}
Muth, R., Galal, T., Heiss, J., Tschorsch, F.: Towards smart contract-based
  verification of anonymous credentials. Cryptology ePrint Archive (2022),
  \url{https://eprint.iacr.org/2022/492}

\bibitem{uPort_naik_2020}
Naik, N., Jenkins, P.: {uPort} open-source identity management system: An
  assessment of self-sovereign identity and user-centric data platform built on
  blockchain. In: International Symposium on Systems Engineering. IEEE (2020)

\bibitem{blogpv_peise_2021}
Peise, M., Kuhlenkamp, J., Busse, A., Eberhardt, J., Ulbricht, M.R., Tai, S.,
  Baus, J., Kassebaum, M., Zörner, T.: Blockchain-based local energy grids:
  Advanced use cases and architectural considerations. In: IEEE 18th
  International Conference on Software Architecture Companion (2021)

\bibitem{ZKlaims_2019}
Schanzenbach, M., Kilian, T., Sch{\"{u}}tte, J., Banse, C.: Zklaims:
  Privacy-preserving attribute-based credentials using non-interactive
  zero-knowledge techniques. In: {ICETE} {(2)}. SciTePress (2019)

\bibitem{zokHealthcare_indian_2020}
Sharma, B., Halder, R., Singh, J.: Blockchain-based interoperable healthcare
  using zero-knowledge proofs and proxy re-encryption. International Conference
  on COMmunication Systems \& NETworkS (COMSNETS)  (2020)

\bibitem{misc:W3_VerifCred}
Sporny, M., Longley, D., Chadwick, D.: Verifiable credentials data model v1.1
  (2021), \url{https://w3.org/TR/vc-data-model/}

\bibitem{credential_revocation_2020}
Vidal, F.R., Gouveia, F., Soares, C.: Revocation mechanisms for academic
  certificates stored on a blockchain. In: 15th Iberian Conference on
  Information Systems and Technologies (CISTI) (2020)

\bibitem{SOK_DeFi_Londoo}
Werner, S.M., Perez, D., Gudgeon, L., Klages{-}Mundt, A., Harz, D.,
  Knottenbelt, W.J.: {SoK}: Decentralized finance ({DeFi}). arXiv  (2021),
  \url{https://arxiv.org/abs/2101.08778}

\bibitem{ethereum:yellowpaper}
Wood, G.: Ethereum: A secure decentralised generalised transaction ledger,
  {B}erlin  (2021), \url{https://github.com/ethereum/yellowpaper/tree/fabef25}

\bibitem{DApps_SoK_2021}
Wu, K., Ma, Y., Huang, G., Liu, X.: A first look at blockchain-based
  decentralized applications. Software: Practice and Experience  (2021)

\end{thebibliography}

\end{document}